# TARGET LOCALIZATION WITH A COPRIME MULTISTATIC MIMO RADAR VIA COUPLED CANONICAL POLYADIC DECOMPOSITION BASED ON JOINT EVD

*Guo-Zhao Liao*[1], *Xiao-Feng Gong*[1], *Wei Liu*[2] *and Hing Cheung So*[3]

[1] School of Information and Communication Engineering, Dalian University of Technology, China
[2] Department of Electrical and Electronic Engineering, The Hong Kong Polytechnic University, Hong Kong, China
[3] Department of Electrical Engineering, City University of Hong Kong, Hong Kong, China

**ABSTRACT**

This paper addresses target localization using a multistatic multiple-input multiple-output (MIMO) radar system with coprime L-shaped receive arrays (CLsA). A target localization method is proposed by modeling the observed signals as tensors that admit a coupled canonical polyadic decomposition (C-CPD) model without matched filtering. It consists of a novel joint eigenvalue decomposition (J-EVD) based (semi-)algebraic algorithm, and a post-processing approach to determine the target locations by fusing the direction-of-arrival estimates extracted from J-EVD-based C-CPD results. Particularly, by leveraging the rotational invariance of Vandermonde structure in CLsA, we convert the C-CPD problem into a J-EVD problem, significantly reducing its computational complexity. Experimental results show that our method outperforms existing tensor-based ones.

***Index Terms***— Multistatic MIMO radar, target localization, coprime L-shaped array, couple canonical polyadic decomposition.

## 1. INTRODUCTION

In the past decade, multistatic (MS) multiple-input multiple-output (MIMO) radar has received increased research focus due to its ability to provide rich spatial diversity and improve target localization accuracy [1–8]. Recently, coupled tensor modeling for MS MIMO radar, which addresses the coupling among arrays and related decomposition techniques, has also gained a lot of attention [1, 5, 7]. From a signal processing perspective, the main step in extending from monostatic/bistatic MIMO radar to MS radar is evolution from single-set techniques to multiset data fusion, as datasets from multiple receive arrays need to be fused. Some methods exploit both array coupling and the multilinear (ML) structure of datasets using structured tensor techniques. For instance, in [1], a structured ML rank-($N_r$, $M_r$, ·) block term decomposition model addresses subarray coupling for MS MIMO radar. In [5], coupled canonical polyadic decomposition (C-CPD) is used without prior waveform knowledge. In [7], a double C-CPD (DC-CPD) model is applied for subarray double coupling. These studies mainly leverage the ML structure of MIMO radar signals without focusing on specific array geometries.

In recent years, array geometries, such as nested, coprime, and minimum redundancy arrays, are widely used for reducing mutual coupling compared to uniform linear arrays (see [9–16] for details). The link between sparse arrays and C-CPD was established in [17–19], leading to a shift from traditional CPD-based methods [20] to coupled tensor-based processing, which connects multidimensional harmonic retrieval (MHR) [21] with multiple invariance ESPRIT [22] to account for harmonic structures. However, previous monostatic-based works have two main limitations: (1) coprime array methods assume uncorrelated sources and many pulses, which may fail with correlated RCS vectors or few pulses; (2) unconstrained C-CPD methods neglect harmonic structures in coprime arrays.

In this study, we propose a target localization method for MS systems employing coprime L-shaped arrays (CLsA) via C-CPD based on joint eigenvalue decomposition (J-EVD). First, we develop a C-CPD model for a MIMO radar system with one transmitter and multiple receivers. Then, by exploiting the rotational invariance property of the Vandermonde structure in CLsA, we transform the C-CPD problem into a J-EVD problem, thus reducing computational complexity. The working conditions for the proposed algorithm are also studied to provide insights into the identifiability of the model. Finally, we employ a post-processing approach to calculate and fuse the DOA information for target localization.

*Notations:* Scalars, vectors, matrices, and tensors are denoted by italic lowercase, lowercase boldface, uppercase boldface, and uppercase calligraphic letters, respectively. The $r$th column and the $(i, j)$ entry of matrix $\mathbf{A}$ are denoted by $\mathbf{a}_r$ and $a_{ij}$, respectively. $|\mathbb{S}|$ denotes the cardinality of a set $\mathbb{S}$. We use MATLAB notations to denote subtensors or submatrices obtained by fixing certain indices or index range of a tensor. For instance, we use $\mathbf{T}_{(:,:,s)}$ to denote the $s$th frontal slice of tensor $\mathcal{T}$ obtained by fixing the third index of $\mathcal{T}$ to $s$. $\mathbf{A}(1:K,:)$ represents the submatrix of $\mathbf{A}$

This work is supported by National Natural Science Foundation of China under grants 62471084, 62071082 and 61871067, and China Postdoctoral Science Foundation under grant 2020M68092.

consisting of the rows from 1 to $K$ of $\mathbf{A}$. Likewise, $\mathbf{A}(\mathbb{Q},:)$ denotes the submatrix of $\mathbf{A}$ consisting of rows indexed by the values of the set $\mathbb{Q}$. We denote the mode-2 matrix representation of a third-order tensor $\mathcal{T} \in \mathbb{C}^{I \times J \times K}$ by $\mathbf{T}_2$. It is defined as: $(\mathbf{T}_2)_{(i-1)K+k,j} = \mathcal{T}_{i,j,k}$. Symbols $(\cdot)^T$, $(\cdot)^{-1}$ and $(\cdot)^\dagger$ denote transpose, inverse and Moore-Penrose pseudo-inverse, respectively; $\circ$ and $\odot$ denote vector outer product and Khatri-Rao product, respectively.

A C-CPD writes a set of tensors $\{\mathcal{T}^{(m)}, m=1,...,M\}$ as the sum of minimal number of coupled rank-1 terms:

$$\mathcal{T}^{(m)} = [\![\mathbf{A}^{(m)}, \mathbf{B}, \mathbf{C}^{(m)}]\!]_R \triangleq \sum_{r=1}^{R} \mathbf{a}_r^{(m)} \circ \mathbf{b}_r \circ \mathbf{c}_r^{(m)}, \quad (1)$$

where $R$ is defined as the coupled rank of $\{\mathcal{T}^{(m)}\}$.

## 2. PROBLEM FORMULATION

Consider an MS MIMO radar system with a single transmit array and multiple receive arrays. All receive arrays are CLsA, while the transmit array is a uniform planar array but could be any configuration. Specifically, the $m$th receive array features two uniform linear subarrays in both x and y-axes. In the x-axis, the first subarray has $I_{x,1}^{(m)}$ sensors with inter-element spacing $M_x d$ and the second $I_{x,2}^{(m)}$ sensors with inter-element spacing $N_x d$, where $M_x$ and $N_x$ are coprime integers, and $d = \lambda/2$, with $\lambda$ being the signal wavelength. Due to their coprime property, these subarrays overlap only at a single reference element. Thus, the total number of x-axis elements is given by $|\mathbb{S}_x^{(m)}| = I_{x,1}^{(m)} + I_{x,2}^{(m)} - 1$. The sensor locations are given by $\mathbb{S}_x^{(m)} d$, where $\mathbb{S}_x^{(m)}$ is an integer set defined as $\mathbb{S}_x^{(m)} = \{(l_{x,i}, 0, 0,) | i = 1, 2, ..., I_{x,1}^{(m)} + I_{x,2}^{(m)} - 1\}$, which can be decomposed as $\mathbb{S}_x^{(m)} = \mathbb{S}_{x,1}^{(m)} \cup \mathbb{S}_{x,2}^{(m)}$, where $\mathbb{S}_{x,1}^{(m)} = \{(M_x m_x, 0, 0,) | 0 \leq m_x \leq I_{x,1}^{(m)} - 1\}$, and $\mathbb{S}_{x,2}^{(m)} = \{(N_x n_x, 0, 0) | 0 \leq n_x \leq I_{x,2}^{(m)} - 1\}$. Here, $I_{x,1}^{(m)} \leq N_x, I_{x,2}^{(m)} \leq M_x$.

Similarly, the y-axis of the $m$th receive array consists of $|\mathbb{S}_y^{(m)}| = I_{y,1}^{(m)} + I_{y,2}^{(m)} - 1$ elements. The sensor locations are given by $\mathbb{S}_y^{(m)} d$, where $\mathbb{S}_y^{(m)}$ is an integer set defined as $\mathbb{S}_y^{(m)} = \{(0, l_{y,i}, 0) | i = 1, 2, ..., I_{y,1}^{(m)} + I_{y,2}^{(m)} - 1\}$, which can be expressed as $\mathbb{S}_y^{(m)} = \mathbb{S}_{y,1}^{(m)} \cup \mathbb{S}_{y,2}^{(m)}$, where $\mathbb{S}_{y,1}^{(m)} = \{(0, M_y m_y, 0) | 0 \leq m_y \leq I_{y,1}^{(m)} - 1\}$, and $\mathbb{S}_{y,2}^{(m)} = \{(0, N_y n_y, 0) | 0 \leq n_y \leq I_{y,2}^{(m)} - 1\}$. Notably, $I_{y,1}^{(m)} \leq N_y$ and $I_{y,2}^{(m)} \leq M_y$.

Consequently, the total number of elements in the $m$th receive array is $|\mathbb{S}^{(m)}| = |\mathbb{S}_x^{(m)} \cup \mathbb{S}_y^{(m)}| = I^{(m)}$, and the sensor locations are given by $\mathbb{S}^{(m)} d$, where $\mathbb{S}^{(m)}$ is an integer set defined as $\mathbb{S}^{(m)} = \mathbb{S}_x^{(m)} \cup \mathbb{S}_y^{(m)}$.

We assume: (A1) the targets are located in the far-field with regard to both transmit and receive arrays; (A2) transmitted and received signals are narrowband; (A3) multiple pulses are emitted in each coherent processing interval (CPI); (A4) The radar cross section (RCS) coefficients of distinct targets vary independently from pulse to pulse (Swerling II model [23]); (A5) DODs and DOAs for different targets are different; (A6) there is no angle ambiguity in each transmit and receive array.

The output signal of the $m$th receive array during the $k$th pulse period can be expressed as (the noise term is omitted for convenience):

$$\mathbf{X}_k^{(m)} = \sum_{r=1}^{R} c_{k,r}^{(m)} \mathbf{a}_r^{(m)} \mathbf{t}_r^T \mathbf{S}^T \in \mathbb{C}^{I^{(m)} \times T}, \quad (2)$$

where $\mathbf{S} \in \mathbb{C}^{T \times J}$ contains probing signals from the transmit array, with each column representing a signal sampled during one pulse period. Here, $T$ is the number of samples per pulse period. $\mathbf{a}_r^{(m)} \in \mathbb{C}^{I^{(m)}}$ and $\mathbf{t}_r \in \mathbb{C}^J$ are the receive and transmit steering vectors, respectively, defined as:

$$\begin{cases} \mathbf{a}_r^{(m)} \triangleq \exp(\mathrm{i} 2\pi \lambda^{-1} [\mathbf{l}_1^{(m)} \mathbf{v}_{(\theta_r^{(m)}, \varphi_r^{(m)})}, ..., \mathbf{l}_{I^{(m)}}^{(m)} \mathbf{v}_{(\theta_r^{(m)}, \varphi_r^{(m)})}]^T), \\ \mathbf{t}_r \triangleq \exp(\mathrm{i} 2\pi \lambda^{-1} [\mathbf{J}_1 \mathbf{v}_{(\alpha_r, \beta_r)}, ..., \mathbf{J}_J \mathbf{v}_{(\alpha_r, \beta_r)}]^T), \end{cases} \quad (3)$$

where 'i' is the imaginary unit, $J$ is the number of antennas in the transmit array, and $I^{(m)}$ is the number in the $m$th receive array; $\mathbf{v}_{(\theta_r^{(m)}, \varphi_r^{(m)})}$ and $\mathbf{v}_{(\alpha_r, \beta_r)}$ denote the DOA of the signal from the $r$th target to the $m$th receive array and the DOD of signal from the transmit array to the $r$th target, respectively. Row vectors $\mathbf{J}_j, \mathbf{l}_i^{(m)} \in \mathbb{R}^3$ represent the sensor locations in the transmit and receive arrays, respectively. Specifically, $\mathbf{l}_i^{(m)} \in \mathbb{S}^{(m)}, i=1,2,...,|\mathbb{S}^{(m)}|$ indicates the position of the $i$th sensor in the coprime array. The transmit array is a uniform planar array. $c_{k,r}^{(m)}$ is the RCS coefficient of the $r$th target relative to the $m$th receive array during the $k$th pulse period, where $r=1,...,R$, $k=1,...,K$, $m=1,...,M$.

Denote $\mathbf{b}_r \triangleq \mathbf{S} \mathbf{t}_r \in \mathbb{C}^T$, and then by (2) we can easily obtain $x_{i,t,k}^{(m)} = \sum_{r=1}^{R} a_{i,r}^{(m)} b_{t,r} c_{k,r}^{(m)}$.

Stacking matrices $\mathbf{X}_k^{(m)}$ for fixed $m$ and varying $k$ along the third mode to construct a third-order tensor $(\mathcal{X}^{(m)})_{:,:,k} \triangleq \mathbf{X}_k^{(m)}$, we have:

$$\mathcal{X}^{(m)} = [\![\mathbf{A}^{(m)}, \mathbf{B}, \mathbf{C}^{(m)}]\!]_R, \quad (4)$$

where $\mathbf{A}^{(m)} \triangleq [\mathbf{a}_1^{(m)}, ..., \mathbf{a}_R^{(m)}]$, $\mathbf{B} \triangleq [\mathbf{b}_1, ..., \mathbf{b}_R]$ and $\mathbf{C}^{(m)} \triangleq [\mathbf{c}_1^{(m)}, ..., \mathbf{c}_R^{(m)}]$. The set of tensors $\{\mathcal{X}^{(m)}, m=1,...,M\}$ together admits a C-CPD.

## 3. PROPOSED METHOD

Now we present a target localization framework for a coprime MS MIMO radar using C-CPD based on J-EVD. The overall process consists of two main stages: the computation of C-CPD based on J-EVD and target localization.

### 3.1. (Semi-)Algebraic C-CPD based on J-EVD

We call an algorithm algebraic if it relies only on arithmetic operations, overdetermined linear equations, singular value decomposition and generalized EVD (GEVD). If some operations are replaced by iterative processes for improved accuracy, the algorithm is called (semi-)algebraic. For the C-CPD of a three-order tensor as shown in (4), we assume that the second factor matrix $\mathbf{B}$ has dimensions $R \times R$, by means of dimensionality reduction [5, 24, 25]. We define $\mathcal{T}^{(m)} \in \mathbb{C}^{I^{(m)} \times R \times K}$ as the result of dimensionality reduction applied to $\mathcal{X}^{(m)}$, where $m=1,...,M$. In the J-EVD-based C-CPD algorithm, the Vandermonde structure in the first factor matrix is used to construct target matrices for J-EVD, yielding an estimate of $\mathbf{B}$ and the Vandermonde generators in factor matrices $\{\mathbf{A}^{(m)}, m=1,...,M\}$.

The proposed algorithm consists of the following steps:

**Step 1: Construct matrices $\mathbf{G}_v^{(m)} \in \mathbb{C}^{R \times R}$ for fixed $m$ and $v=1,...,4$**

To start, we define the following notations. For the $m$th receive array using CLsA, the steering vectors are represented by the matrix $\mathbf{A}_x^{(m)}$ of dimensions $|\mathbb{S}_x^{(m)}| \times R$, given by $\mathbf{A}_x^{(m)} = \mathbf{A}^{(m)}(\mathbb{Q}_x^{(m)},:)$, and the matrix $\mathbf{A}_y^{(m)}$ of dimensions $|\mathbb{S}_y^{(m)}| \times R$, given by $\mathbf{A}_y^{(m)} = \mathbf{A}^{(m)}(\mathbb{Q}_y^{(m)},:)$. Here, $\mathbb{Q}_x^{(m)}$ and $\mathbb{Q}_y^{(m)}$ are index sets corresponding to $\mathbb{S}_x^{(m)}$ and $\mathbb{S}_y^{(m)}$ within the set $\mathbb{S}^{(m)}$, respectively. The matrices $\mathbf{A}_{x,1}^{(m)}$ and $\mathbf{A}_{x,2}^{(m)}$ are the Vandermonde matrices for the two uniform linear subarrays along the x-axis, defined as $\mathbf{A}_{x,1}^{(m)} = \mathbf{A}_x^{(m)}(\mathbb{Q}_{x,1}^{(m)},:)$ and $\mathbf{A}_{x,2}^{(m)} = \mathbf{A}_x^{(m)}(\mathbb{Q}_{x,2}^{(m)},:)$, respectively. Here, $\mathbb{Q}_{x,1}^{(m)}$ and $\mathbb{Q}_{x,2}^{(m)}$ are index sets corresponding to $\mathbb{S}_{x,1}^{(m)}$ and $\mathbb{S}_{x,2}^{(m)}$ within the set $\mathbb{S}_x^{(m)}$. Similarly, for the y-axis, the Vandermonde matrices are $\mathbf{A}_{y,1}^{(m)} = \mathbf{A}_y^{(m)}(\mathbb{Q}_{y,1}^{(m)},:)$ and $\mathbf{A}_{y,2}^{(m)} = \mathbf{A}_y^{(m)}(\mathbb{Q}_{y,2}^{(m)},:)$, where $\mathbb{Q}_{y,1}^{(m)}$ and $\mathbb{Q}_{y,2}^{(m)}$ are index sets of $\mathbb{S}_{y,1}^{(m)}$ and $\mathbb{S}_{y,2}^{(m)}$ within $\mathbb{S}_y^{(m)}$, respectively. Next, we explain how $\mathbf{G}_1^{(m)}$ is constructed by leveraging the Vandermonde structure present in $\mathbf{A}_{x,1}^{(m)}$. First, we need to extract the data tensor corresponding to the x-axis sensor elements of the $m$th receive array, which can be expressed as follows:

$$\mathcal{T}_x^{(m)} = \mathcal{T}^{(m)}(\mathbb{Q}_x^{(m)},:,:) = [\![\mathbf{A}_x^{(m)}, \mathbf{B}, \mathbf{C}^{(m)}]\!]_R. \quad (5)$$

Then, we extract the data tensor corresponding to the first uniform linear array along the x-axis:

$$\mathcal{T}_{x,1}^{(m)} = \mathcal{T}_x^{(m)}(\mathbb{Q}_{x,1}^{(m)},:,:) = [\![\mathbf{A}_{x,1}^{(m)}, \mathbf{B}, \mathbf{C}^{(m)}]\!]_R. \quad (6)$$

After that, since $\mathcal{T}_{x,1}^{(m)}$ admits a CPD (6), its mode-2 matrix representation $\mathbf{T}_{x,1}^{(m)}$ is:

$$\mathbf{T}_{x,1}^{(m)} = (\mathbf{A}_{x,1}^{(m)} \odot \mathbf{C}^{(m)}) \cdot \mathbf{B}^T. \quad (7)$$

We choose two submatrices $\mathbf{M}_{x,1}^{(m,1)}$ and $\mathbf{M}_{x,1}^{(m,2)}$ from $\mathbf{T}_{x,1}^{(m)}$, each having dimensions $(I_{x,1}^{(m)}-1)K \times R$:

$$\begin{cases} \mathbf{M}_{x,1}^{(m,1)} \triangleq \mathbf{T}_{x,1}^{(m)}(1:(I_{x,1}^{(m)}-1)K,:) = (\underline{\mathbf{A}}_{x,1}^{(m)} \odot \mathbf{C}^{(m)}) \cdot \mathbf{B}^T, \\ \mathbf{M}_{x,1}^{(m,2)} \triangleq \mathbf{T}_{x,1}^{(m)}(K+1:I_{x,1}^{(m)}K,:) = (\overline{\mathbf{A}}_{x,1}^{(m)} \odot \mathbf{C}^{(m)}) \cdot \mathbf{B}^T, \end{cases} \quad (8)$$

where $\underline{\mathbf{A}}^{(1)}, \overline{\mathbf{A}}^{(1)} \in \mathbb{C}^{(I_1-1) \times R}$ and $\overline{\mathbf{A}}^{(1)} \in \mathbb{C}^{(I_1-1) \times R}$ are obtained by removing the last and first rows of $\mathbf{A}^{(1)}$, respectively. As $\mathbf{A}_{x,1}^{(m)}$ is a Vandermonde matrix, we have $\overline{\mathbf{A}}_{x,1}^{(m)} = \underline{\mathbf{A}}_{x,1}^{(m)} \cdot \mathbf{Z}_{x,1}^{(m)}$, where $\mathbf{Z}_{x,1}^{(m)}$ is a diagonal matrix holding the Vandermonde generators of $\mathbf{A}_{x,1}^{(m)}$, $z_{x1,1}^{(m)},...,z_{x1,R}^{(m)}$, in its main diagonal. Therefore, we have the following result:

$$\overline{\mathbf{A}}_{x,1}^{(m)} \odot \mathbf{C}^{(m)} = (\underline{\mathbf{A}}_{x,1}^{(m)} \odot \mathbf{C}^{(m)}) \cdot \mathbf{Z}_{x,1}^{(m)}. \quad (9)$$

Assume that $\mathbf{M}_{x,1}^{(m,1)}$ has full column rank, and construct the target matrix as $\mathbf{G}_1^{(m)} \triangleq [(\mathbf{M}_{x,1}^{(m,1)})^\dagger \mathbf{M}_{x,1}^{(m,2)}]^T \in \mathbb{C}^{R \times R}$. Then, after simple derivations, we obtain:

$$\mathbf{G}_1^{(m)} \triangleq [(\mathbf{M}_{x,1}^{(m,1)})^\dagger \mathbf{M}_{x,1}^{(m,2)}]^T = \mathbf{B} \cdot \mathbf{Z}_{x,1}^{(m)} \cdot \mathbf{B}^{-1} \quad (10)$$

Next, we follow similar procedures to exploit the Vandermonde structure in $\mathbf{A}_{x,2}^{(m)}$, $\mathbf{A}_{y,1}^{(m)}$ and $\mathbf{A}_{y,2}^{(m)}$ to construct matrices $\mathbf{G}_2^{(m)}, \mathbf{G}_3^{(m)}$ and $\mathbf{G}_4^{(m)}$, respectively. For instance, for the construction of $\mathbf{G}_2^{(m)}$, we replace $\mathbb{Q}_{x,1}^{(m)}$ in (6) with $\mathbb{Q}_{x,2}^{(m)}$ to obtain the data tensor $\mathcal{T}_{x,2}^{(m)}$ corresponding to the x-axis direction of the $m$th receive array. Then, $\mathbf{T}_{x,2}^{(m)}$ is obtained by (7). The submatrices of $\mathbf{T}_{x,2}^{(m)}$, namely $\mathbf{M}_{x,2}^{(m,1)}$ and $\mathbf{M}_{x,2}^{(m,2)}$, are constructed by (8). Subsequently, $\mathbf{G}_2^{(m)}$ is computed with $\mathbf{M}_{x,2}^{(m,1)}$ and $\mathbf{M}_{x,2}^{(m,2)}$ by (10).

Analogously, we can construct $\mathbf{G}_3^{(m)}$ and $\mathbf{G}_4^{(m)}$ using the sets $\mathbb{Q}_{y,1}^{(m)}$ and $\mathbb{Q}_{y,2}^{(m)}$ such that $\mathbf{G}_1^{(m)}, \mathbf{G}_2^{(m)}, \mathbf{G}_3^{(m)}$ and $\mathbf{G}_4^{(m)}$ together admit the following J-EVD formulation:

$$\begin{cases} \mathbf{G}_1^{(m)} = \mathbf{B} \cdot \mathbf{Z}_{x,1}^{(m)} \cdot \mathbf{B}^{-1}, \mathbf{G}_2^{(m)} = \mathbf{B} \cdot \mathbf{Z}_{x,2}^{(m)} \cdot \mathbf{B}^{-1}, \\ \mathbf{G}_3^{(m)} = \mathbf{B} \cdot \mathbf{Z}_{y,1}^{(m)} \cdot \mathbf{B}^{-1}, \mathbf{G}_4^{(m)} = \mathbf{B} \cdot \mathbf{Z}_{y,2}^{(m)} \cdot \mathbf{B}^{-1}, \end{cases} \quad (11)$$

where $\mathbf{Z}_{x,1}^{(m)}, \mathbf{Z}_{x,2}^{(m)}, \mathbf{Z}_{y,1}^{(m)}$ and $\mathbf{Z}_{y,2}^{(m)}$ are diagonal, holding in their main diagonals the Vandermonde generators of $\mathbf{A}_{x,1}^{(m)}$, $\mathbf{A}_{x,2}^{(m)}, \mathbf{A}_{y,1}^{(m)}$ and $\mathbf{A}_{y,2}^{(m)}$, respectively. Note that in this study, J-EVD aims to find an invertible matrix $\mathbf{B}$, such that $\mathbf{B}^{-1} \cdot \mathbf{G}_v^{(m)} \cdot \mathbf{B}$ is diagonal for $v \in \{1,2,3,4\}$.

Note that at least one of matrices $\mathbf{M}_{x,1}^{(m,1)}, \mathbf{M}_{x,2}^{(m,1)}, \mathbf{M}_{y,1}^{(m,1)}$ and $\mathbf{M}_{y,2}^{(m,1)}$ for $m=1,...,M$, is required to have full column rank. Generically[1], this condition translates to the following working conditions for the proposed J-EVD-based C-CPD algorithm:

$$\begin{cases} \min(T,J) \geq R, \\ (I'-1)K \geq R, \end{cases} \quad (12)$$

where $I'$ represents the value corresponding to the maximum element in the set $\{I_{x,1}^{(m)}, I_{x,2}^{(m)}, I_{y,1}^{(m)}, I_{y,2}^{(m)}, m=1,...,M\}$.

**Step 2: Compute B via J-EVD of $\{\mathbf{G}_v^{(m)}, v=1,...,4, m=1,...,M\}$, and $\mathbf{A}^{(m)}$, $\mathbf{C}^{(m)}$ using rank-1 approximation**

This algorithm converts the J-EVD problem as a structured CPD problem:

$$\mathcal{G} = [\![\mathbf{B}, \mathbf{D}, \mathbf{F}]\!]_R, \quad \text{s.t., } \mathbf{B} \cdot \mathbf{D}^T = \mathbf{I}, \quad (13)$$

where

$$\begin{cases} \mathcal{G}_{(:,:,w)} \triangleq \mathbf{G}_w', w \in \{1,2,3,..,4M\}, \\ \mathbf{F} \triangleq [\mathrm{diag}(\mathbf{Z}^{(1)}), \mathrm{diag}(\mathbf{Z}^{(2)}),...,\mathrm{diag}(\mathbf{Z}^{(M)})]^T, \end{cases} \quad (14)$$

with $\mathbf{Z}^{(m)} = [\mathrm{diag}(\mathbf{Z}_{x,1}^{(m)}), \mathrm{diag}(\mathbf{Z}_{x,2}^{(m)}), \mathrm{diag}(\mathbf{Z}_{y,1}^{(m)}), \mathrm{diag}(\mathbf{Z}_{y,2}^{(m)})]$ and $\mathbf{G}_{4(m-1)+v}' = \mathbf{G}_v^{(m)}, v=1,...4, m=1,...,M$.

We can obtain the factor matrix $\mathbf{B}$ after solving the CPD of $\mathcal{G}$ formed from the J-EVD problem (11) for all $m$. Since we assume that matrix $\mathbf{B}$ has full column rank, the tensor $\mathcal{G}$ has dimensions $R \times R \times 4M$, where $4M \geq 2$. Consequently, the CPD problem (13) can be addressed algebraically through GEVD. We use Tensorlab+ [28] to perform both GEVD and simultaneous generalized Schur decomposition (SGSD) algorithms, with GEVD supplying initial values to SGSD to refine $\mathbf{B}$ for improved accuracy.

---

[1] A property is called generic if it holds with probability one in the Lebesgue measure.

Once **B** is computed, we construct $\Omega_r^{(m)} \triangleq \text{unvec}((\mathbf{T}_2^{(m)} \mathbf{B}^{-T})_{(:,r)})$ to calculate the factor matrices $\mathbf{A}^{(m)}$ and $\mathbf{C}^{(m)}$. We than approximate $\Omega_r^{(m)}$ by a rank-1 matrix: $\Omega_r^{(m)} = \tilde{\mathbf{a}}_r^{(m)} \tilde{\mathbf{c}}_r^{(m)T}$, where $\mathbf{a}_r^{(m)}$ and $\mathbf{c}_r^{(m)}$ are the dominant left and right singular vectors, respectively.

The computational complexity of the algebraic C-CPD-based J-EVD algorithm is $O(3(I_{x,1}^{(m)}-1)KR^2)$ for target matrix construction, and $O(30R^2)$ per pair of $\mathcal{G}(:,:,w_1)$ and $\mathcal{G}(:,:,w_2)$, $w_1 \neq w_2$, for solving the CPD problem via GEVD, as in (13).

### 3.2. Target Localization

We use estimates of the first matrices, i.e., receive steering vectors, to compute the locations of the targets via two steps:

(i) After applying the (semi-)algebraic C-CPD algorithm based on J-EVD, factor matrices $\tilde{\mathbf{A}}^{(m)}$ for all receiving arrays are obtained for $m=1,...,M$. The steering vectors along the x-axis and y-axis for the *m*th array are extracted as $\tilde{\mathbf{A}}_x^{(m)} = \tilde{\mathbf{A}}^{(m)}(\mathbb{Q}_x^{(m)},:)$ and $\tilde{\mathbf{A}}_y^{(m)} = \tilde{\mathbf{A}}^{(m)}(\mathbb{Q}_y^{(m)},:)$, respectively. These are used to form a virtual coprime planar array steering matrix $\breve{\mathbf{A}}^{(m)} = \tilde{\mathbf{A}}_x^{(m)} \odot \tilde{\mathbf{A}}_y^{(m)}$. DOA parameters for each target can be computed from $\breve{\mathbf{A}}^{(m)}$ using the single-source MHR method [17, 19].

(ii) For each target, fuse DOA estimates from various receive arrays to determine the target's position as described in [5]. More precisely, let $\xi \triangleq [\xi_x, \xi_y, \xi_z]^T$ be a point in space, and $\mathbf{p}_r^{(m)} \in \mathbb{R}^3$ be the location of the center of the *m*th receive array. The squared distance between $\xi$ and the line through $\mathbf{p}_r^{(m)}$ along DOA $\mathbf{v}_{(\bar{\theta}_r^{(m)}, \bar{\varphi}_r^{(m)})}$ is given by $d_r^{(m)2} = \|\xi - \mathbf{p}_r^{(m)}\|^2 - \|(\xi - \mathbf{p}_r^{(m)})^T \mathbf{v}_{(\bar{\theta}_r^{(m)}, \bar{\varphi}_r^{(m)})}\|^2$. The target location is the point that minimizes the sum of these squared distances, i.e., the point closest to all estimated paths defined by the receive array positions and associated DOAs:

$$\tilde{\mathbf{z}}_r = \arg\min_\xi (\sum_{m=1}^M d_r^{(m)2}). \tag{15}$$

## 4. SIMULATION RESULTS

We present simulation results to evaluate the proposed method and compare it with existing tensor-based methods, including C-CPD-MHR-SD [18], which is based on MHR and C-CPD, and CPD-NLS(ALG) [28], a nonlinear least squares (NLS) based CPD method initialized with the CPD algebraic algorithm [26]. The mean angular error (MAE), root mean square error (RMSE) and average CPU time are used as performance metrics. We refer to [5] and [27] for the definitions of MAE and RMSE, respectively.

The MIMO radar consists of one transmit array and three receive arrays with $I \triangleq I^{(1)} = I^{(2)} = I^{(3)}$. The transmit array is located at (0, –8000λ, 0). The receive arrays are CLsA, located at (–8000λ, 8000λ, 0), (0, 8000λ, 0), and (8000λ, 8000λ, 0), respectively. The target locations are randomly drawn in the area with x-coordinate ranging from –7000λ to 7000λ, y-coordinate ranging from –7000λ to 7000λ and z-coordinate ranging from 4000λ to 8000λ, respectively.

We mainly consider two cases: the overdetermined case with both $I$ and $K$ higher than $R$, and the underdetermined case with $I$ and $K$ smaller than $R$. In both Case 1 and Case 2, all CLsA have the same coprime configuration. More precisely, $M_x = M_y = 4, N_x = N_y = 7, I_{x,1}^{(m)} = I_{x,2}^{(m)} = I_{y,1}^{(m)} = I_{y,2}^{(m)} = 4$, and $I = 13$. Figs. 1 and 2 present the MAE curves, RMSE curves, and average CPU time curves versus signal-to-noise ratio (SNR) based on averages of 200 Monte Carlo runs. The Cramér-Rao bound (CRB) based lower bound curves for MAE and RMSE, labeled as CRB-MAE and CRB-RMSE, as defined in [5] and [27].

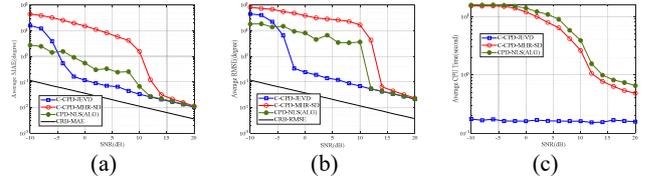

**Fig. 1**. (a) MAE, (b) RMSE, and (c) CPU time vs. SNR in Case 1.

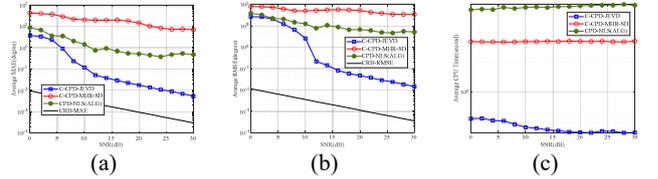

**Fig. 2**. (a) MAE, (b) RMSE, and (c) CPU time vs. SNR in Case 2.

In Case 1 (*I*=13, *J*=16, *K*=15, *T*=64, *R*=10), as shown in Fig. 1, simultaneous diagonalization (SD) based methods such as C-CPD-MHR-SD perform well only at high SNRs (12–20dB) due to noise sensitivity. The proposed (semi-)algebraic C-CPD algorithm based on J-EVD method provides robust performance at lower SNR levels (-6–10dB) and significantly outperforms NLS-based CPD algorithms.

In Case 2 (*I*=13, *J*=49, *K*=20, *T*=64, *R*=25), as depicted in Fig. 2, the C-CPD-JEVD method is the only one producing effective results, as SD-based methods are too sensitive to noise and CPD-NLS(ALG) does not fully utilize the underlying Vandermonde structures in coprime array.

Those two cases show that the proposed method outperforms C-CPD-MHR-SD and CPD-NLS(ALG) in both overdetermined and underdetermined cases in terms of accuracy and computational cost.

## 5. CONCLUSION

A target localization method has been proposed for MS systems employing CLsA via C-CPD based on J-EVD. An algebraic C-CPD algorithm based on J-EVD is introduced by exploiting the Vandermonde structure in CLsA, along with a post-processing approach for calculating target locations by fusing DOA parameters extracted from the J-EVD based C-CPD results. The working conditions for the proposed algorithm are studied to provide insights into the identifiability of the model. Experimental results demonstrate that our method outperforms existing tensor-based approaches.